\documentclass[12pt,epsf]{article}
\usepackage{graphicx}
\usepackage{epsfig}
\begin{document}

\def\a{\alpha}
\def\b{\beta}
\def\c{\varepsilon}
\def\d{\delta}
\def\e{\epsilon}
\def\f{\phi}
\def\g{\gamma}
\def\h{\theta}
\def\k{\kappa}
\def\l{\lambda}
\def\m{\mu}
\def\n{\nu}
\def\p{\psi}
\def\q{\partial}
\def\r{\rho}
\def\s{\sigma}
\def\t{\tau}
\def\u{\upsilon}
\def\v{\varphi}
\def\w{\omega}
\def\x{\xi}
\def\y{\eta}
\def\z{\zeta}
\def\D{\Delta}
\def\G{\Gamma}
\def\H{\Theta}
\def\L{\Lambda}
\def\F{\Phi}
\def\P{\Psi}
\def\S{\Sigma}

\def\o{\over}
\def\beq{\begin{eqnarray}}
\def\eeq{\end{eqnarray}}
\newcommand{\gsim}{ \mathop{}_{\textstyle \sim}^{\textstyle >} }
\newcommand{\lsim}{ \mathop{}_{\textstyle \sim}^{\textstyle <} }
\newcommand{\vev}[1]{ \left\langle {#1} \right\rangle }
\newcommand{\bra}[1]{ \langle {#1} | }
\newcommand{\ket}[1]{ | {#1} \rangle }
\newcommand{\EV}{ {\rm eV} }
\newcommand{\KEV}{ {\rm keV} }
\newcommand{\MEV}{ {\rm MeV} }
\newcommand{\GEV}{ {\rm GeV} }
\newcommand{\TEV}{ {\rm TeV} }
\def\diag{\mathop{\rm diag}\nolimits}
\def\Spin{\mathop{\rm Spin}}
\def\SO{\mathop{\rm SO}}
\def\O{\mathop{\rm O}}
\def\SU{\mathop{\rm SU}}
\def\U{\mathop{\rm U}}
\def\Sp{\mathop{\rm Sp}}
\def\SL{\mathop{\rm SL}}
\def\tr{\mathop{\rm tr}}

\def\IJMP{Int.~J.~Mod.~Phys. }
\def\MPL{Mod.~Phys.~Lett. }
\def\NP{Nucl.~Phys. }
\def\PL{Phys.~Lett. }
\def\PR{Phys.~Rev. }
\def\PRL{Phys.~Rev.~Lett. }
\def\PTP{Prog.~Theor.~Phys. }
\def\ZP{Z.~Phys. }

\newcommand{\bear}{\begin{array}}  
\newcommand {\eear}{\end{array}}
\newcommand{\la}{\left\langle}  
\newcommand{\ra}{\right\rangle}
\newcommand{\non}{\nonumber}  
\newcommand{\ds}{\displaystyle}
\newcommand{\red}{\textcolor{red}}
\def\ubl{U(1)$_{\rm B-L}$}
\def\REF#1{(\ref{#1})}
\def\lrf#1#2{ \left(\frac{#1}{#2}\right)}
\def\lrfp#1#2#3{ \left(\frac{#1}{#2} \right)^{#3}}
\def\OG#1{ {\cal O}(#1){\rm\,GeV}}


\baselineskip 0.7cm

\begin{titlepage}

\begin{flushright}
UT-09-07\\
IPMU 09-0027
\end{flushright}

\vskip 1.35cm
\begin{center}
{\large \bf
    Decaying Hidden Gaugino as a Source of PAMELA/ATIC Anomalies
}
\vskip 1.2cm
Satoshi Shirai$^1$, Fuminobu Takahashi$^2$ and  T. T. Yanagida$^{1,2}$
\vskip 0.4cm

{\it $^1$  Department of Physics, University of Tokyo,\\
     Tokyo 113-0033, Japan\\
$^2$ Institute for the Physics and Mathematics of the Universe, 
University of Tokyo,\\ Chiba 277-8568, Japan}

\vskip 1.5cm

\abstract{ We study a scenario that a U(1) hidden gaugino constitutes
  the dark matter in the Universe and decays into a lepton and slepton
  pair through a mixing with a U(1)$_{\rm B-L}$ gaugino.  We find that
  the dark-matter decay can account for the recent PAMELA and ATIC
  anomalies in the cosmic-ray positrons and electrons without an
  overproduction of antiprotons.  }
\end{center}
\end{titlepage}

\setcounter{page}{2}

\section{Introduction}
\label{sec:1}

In the string landscape of vacua, the presence of many gauge
symmetries besides the standard-model (SM) gauge groups is a quite
common phenomenon. Therefore, it is very interesting to consider not
only low-energy supersymmetry (SUSY) but also additional low-energy
gauge symmetries.

If some of the extra U(1)'s are unbroken, the corresponding gauginos
may receive SUSY-breaking soft masses of the order of the gravitino
mass, $m_{3/2}\simeq O(1)$\,TeV. If those U(1)'s are confined on one
brane far separated from another brane on which the SUSY SM (SSM) particles live,
direct couplings between the U(1) gauge multiplets and the SSM
particles would be exponentially suppressed~\cite{Chen:2008md}. Then
the gauginos of the extra U(1)'s may have lifetimes much longer than
the age of the Universe, and therefore can be candidates for dark
matter (DM) in the Universe.  We call such gauginos as hidden gauginos
and consider a scenario that they constitute the
DM~\cite{Ibarra:2008kn}.

The purpose of this paper is to show that the decays of the hidden
gauginos naturally explain anomalous excesses in the cosmic-ray
electron/positron fluxes observed by PAMELA~\cite{Adriani:2008zr} and
ATIC~\cite{:2008zz}.  As we will see in the following sections, the
decay proceeds through a mixing with the \ubl \,\,gaugino.\footnote{
See Refs.~\cite{Chen:2008yi,Chen:2008md} for the non-supersymmetric counterpart. 
} The hidden gaugino decays into a lepton and slepton pair, and the
slepton will further decay into a lepton and a neutralino.  Those
energetic leptons from the DM decay are the source of the PAMELA/ATIC
anomalies.

The PAMELA experiment also reported the antiproton/proton ratio, which
provides a severe constraint on the DM contribution to the antiproton
flux~\cite{Adriani:2008zq}.  It is quite remarkable that the present
model is free from an overproduction of the antiproton, if the decay
into a quark and squark pair is kinematically
forbidden~\cite{Arvanitaki:2008hq}. (See
Refs.~\cite{Cholis:2008vb,Chen:2008dh,Chen:2008md,Hamaguchi:2008ta,Ibe:2009dx} for
other ways to avoid the problem of the antiprotons.) Our model
predicts a bump at several hundred GeV in the diffuse gamma-ray flux,
which may be tested by Fermi Gamma-ray Space Telescope~\cite{FGST} in
operation. Also, since the mass scale of the SSM particles
is not directly related to the energy scale of the electron/positron
excesses, the gluino mass can be lighter than e.g. $1$\,TeV, and hence
SUSY may be discovered at LHC.

\section{Model and decay of hidden gaugino}
\label{sec:2}

To demonstrate our point we consider a reduced model in $(4+1)$
dimensional space time. The extra dimension is assumed to be
compactified on $S^1/Z_2$ which has two distinct boundaries.  Suppose
that a hidden U(1) gauge multiplet $(\lambda_{H},A_{H}, D_{H})$ is
confined on one boundary and the SSM multiplets on the other. In such
a set-up, direct interactions between the two sectors are suppressed
by a factor of $\exp({- M_* L})$, where $M_*$ is the five-dimensional
Planck scale and $L$ denotes the size of the extra dimension. For
e.g. $M_* L \sim 10^2$, the direct couplings are so suppressed that
the hidden gaugino will be practically stable in a cosmological time
scale~\cite{Chen:2008md}.\footnote{
With our choice of $L \sim 10^2/M_*$, the five dimensional Planck scale 
$M_*$ is roughly equal to $M_P /10 \simeq 10^{17}$ GeV
which is larger than the grand unification theory (GUT) scale ($\sim 10^{16}$ GeV), 
and the subsequent analysis in the text is valid.
}

We introduce a U(1)$_{\rm B-L}$ gauge multiplet $(\lambda_{\rm
  B-L},A_{\rm B-L}, D_{\rm B-L})$ and a SUSY breaking multiplet $Z$ in
the bulk. The multiplet $Z$ is assumed to have a SUSY-breaking F-term,
$|F_Z| = \sqrt{3} m_{3/2} M_P$, where $M_P \simeq 2.4 \times 10^{18}$
GeV. As we will see below, the presence of the U(1)$_{\rm B-L}$ in the
bulk enables the hidden U(1) gaugino to decay into the SSM particles
through an unsuppressed kinetic mixing between the two U(1)'s.

In order to ensure the anomaly cancellation of U(1)$_{\rm B-L}$, it is
necessary to introduce three generations of right-handed neutrinos
$N$.  The seesaw mechanism~\cite{seesaw} for neutrino mass generation
suggests the Majorana mass of the (heaviest) right-handed neutrino at
about the GUT scale. Such a large Majorana mass can be naturally
provided if the U(1)$_{\rm B-L}$ symmetry is spontaneously broken at a
scale around $10^{16}$ GeV.  To be explicit, let us introduce two
supermultiplets $\Phi(2)$ and $\bar{\Phi}(-2)$, where the $B-L$ charges
are shown in the parentheses.  In order to induce the $B-L$ breaking,
we consider the superpotential,
\beq
W\;=\; X (\Phi \bar \Phi - v_{\rm B-L}^2),
\eeq
where $X$ is a gauge-singlet multiplet and $v_{\rm B-L}$ represents
the $B-L$ breaking scale.  In the vacuum $\Phi$ and $\bar \Phi$
acquire non-vanishing expectation values: $\la \Phi \ra = \langle \bar
\Phi \rangle = v_{\rm B-L}$, and the U(1)$_{\rm B-L}$ is spontaneously
broken.  As a result, the gauge boson as well as the gaugino acquire a
mass of $M \equiv 4 g_{\rm B-L} v_{\rm B-L}$, where $g_{\rm B-L}$
denotes the U(1)$_{\rm B-L}$ gauge coupling.  In particular, the
gaugino $\lambda_{\rm B-L}$ forms a Dirac mass term with $\Psi \equiv
(\chi_\Phi - \chi_{\bar \Phi})/\sqrt{2}$, where $\chi_\Phi$ and
$\chi_{\bar \Phi}$ are the fermionic components of $\Phi$ and $\bar
\Phi$, respectively.

Let us estimate the mixing between the hidden U(1) and U(1)$_{\rm
  B-L}$ multiplets.  The kinetic terms are given by
\beq
{\cal L }_K &=& \frac{1}{4}\int d^2\theta (W_{ H}W_{H} + W_{\rm B-L}W_{\rm B-L} + 2 \kappa W_{H}W_{\rm B-L})  
+ {\rm h.c.}, \non\\
&\supset&-i \left( \bar \lambda_{H} \bar \sigma^\mu \partial_\mu \lambda_{ H}
+ \bar \lambda_{\rm B-L} \bar \sigma^\mu \partial_\mu \lambda_{\rm B-L}
+  \kappa \bar \lambda_{H} \bar \sigma^\mu \partial_\mu \lambda_{\rm B-L}
+  \kappa \bar \lambda_{\rm B-L} \bar \sigma^\mu \partial_\mu \lambda_{H}
\right),
\label{eq:mixing}
\eeq
where $\kappa$ is a kinetic mixing parameter of $O(0.1)$, and we have
extracted relevant terms in the second equality.  On the other hand,
the mass terms are given by
\beq
{\cal L}_M \;=\; -\frac{1}{2}m \lambda_H \lambda_H - M \lambda_{\rm B-L}\Psi + {\rm h.c.},
\label{eq:mass}
\eeq
where we have assumed that the hidden gaugino $\lambda_H$ acquires a
Majorana mass $m$ through the SUSY breaking effect.\footnote{
The $\lambda_{\rm B-L}$ may also acquire a Majorana mass of $O(m)$
from the SUSY breaking effect. Including the mass does not change
the following arguments.
} (Note that we assume that the hidden U(1) remains unbroken in the low
energy.) We assume that $m$ is of the order of the weak scale, while $M$ is around the
GUT scale $\sim 10^{16}$ GeV.
The canonically normalized mass eigenstates
$(\lambda_1,\lambda_2,\lambda_3)$ have masses $\approx m,
M/\sqrt{1-\kappa^2}, -M/\sqrt{1-\kappa^2}$. They are related with the
gauge eigenstates $(\lambda_{H}, \lambda_{\rm B-L},\Psi)$
as,\footnote{
A similar  mixing could arise from an interaction $\int d^2 \theta \,Z W_{
    H} W_{\rm B-L}/M_P$.  For $m_{3/2} \sim m$, the resultant mixing
  angle is of the same order of magnitude as that in the text.
 }
\beq
\left(
\bear{c}
\lambda_1\\
\lambda_2\\
\lambda_3
\eear
\right)
 \;\simeq\;
\left(
\bear{ccc}
1& {\kappa}& { \frac{ \kappa m}{M}}\\
{- \frac{\kappa m}{\sqrt{2} M}}&{\frac{1}{\sqrt{2}}}&{\frac{1}{\sqrt{2}}}\\
{-\frac{\kappa m}{\sqrt{2} M}}&{-\frac{1}{\sqrt{2}}}&{\frac{1}{\sqrt{2}}}
\eear
\right)
\left(
\bear{c}
\lambda_{H}\\
\lambda_{\rm B-L}\\
\Psi
\eear
\right),
\label{eq:rel1}
\eeq
or equivalently,
\beq
\left(
\bear{c}
\lambda_{H}\\
\lambda_{\rm B-L}\\
\Psi
\eear
\right)
 \;\simeq\;
\left(
\bear{ccc}
1& {-\frac{\kappa}{\sqrt{2}}}& { \frac{\kappa}{\sqrt{2}}}\\
{ \frac{\kappa m^2}{M^2}}&{\frac{1}{\sqrt{2}}}&{-\frac{1}{\sqrt{2}}}\\
{ \frac{ \kappa m}{M}}&{\frac{1}{\sqrt{2}}}&{\frac{1}{\sqrt{2}}}
\eear
\right)
\left(
\bear{c}
\lambda_1\\
\lambda_2\\
\lambda_3
\eear
\right),
\label{eq:rel2}
\eeq
where we have approximated $m \ll M$ and $\kappa \lsim 0.1$.  The
lightest fermion $\lambda_1$ is almost the same state as $\lambda_H$ and is
the candidate for the DM.

Using the relation \REF{eq:rel2} we can derive the interaction between
$\lambda_1$ and the SSM particles.  The $\lambda_{\rm B-L}$ has
interactions with the SSM particles via the U(1)$_{\rm B-L}$ gauge
symmetry;
\beq
-\sqrt{2}g_{\rm B-L} Y_{\psi}  \lambda_{\rm B-L} \phi^{*}_{\rm SSM} \psi_{\rm SSM}  + {\rm h.c.}, 
\eeq
where $\psi$ is the SM fermion, $\phi$ is its scalar partner and
$Y_{\psi}$ is their (B$-$L) number.  Through the mixing shown in
Eq.~(\ref{eq:rel2}), the $\lambda_1$ gets the interaction with the SSM
particles as
\beq
{\cal L}_{\rm int}\;=\;
-\sqrt{2}g_{\rm B-L} Y_{\psi} \kappa \lrfp{m}{M}{2} \lambda_{1} \phi^{*}_{\rm SSM} \psi_{\rm SSM}  + {\rm h.c.}, 
 \label{eq:int}
 \eeq%
 which enables the $\lambda_1$ to decay into SSM particles.  Its
 lifetime is estimated  to be
\beq
\Gamma_{\rm DM}^{-1} ({\rm DM} \to \phi+\psi)\simeq 
2\times 10^{26}~{\rm sec}~  g^{-2}_{\rm B-L} Y^{-2}_{\psi}\kappa^{-2} \left( 1- \frac{m_{\phi}}{m_{\rm DM}}\right)^{-2}
\left(\frac{m}{1~{\rm TeV}} \right)^{-5} \left(\frac{M}{10^{16} ~\GEV}\right)^{4}\frac{1}{C_{\psi}},
\eeq
where $C_{\psi}$ is a color factor of $\psi$, i.e., $3$ for quarks and $1$ for leptons.  It is quite remarkable
that the hierarchy between the $B-L$ breaking scale $\sim 10^{16}$ GeV
and the SUSY breaking mass of the hidden gaugino of $O(1)$\,TeV
naturally leads to the lifetime of $O(10^{26})$ second that is
needed to account for the positron excess.

From Eq.~(\ref{eq:rel2}) we can see that the mixing of $\lambda_1$
with $\Psi$ is much larger than that with $\lambda_{\rm B-L}$. Thus we
need to make sure that the coupling of $\Psi$ to the SSM particles
must be small enough. The most dangerous coupling is that with the
Higgs multiplets,
\beq
W\;=\; \frac{c}{M_{*}} \Phi \bar{\Phi} H_1 H_2,
\eeq
where $c$ is a numerical coefficient.  If $c$ is of order unity, this
operator induces a too fast decay of $\lambda_1$ to become DM, since
$\Phi$ and $\bar{\Phi}$ acquire large expectation values, $\langle
\Phi \rangle = \langle \bar{\Phi} \rangle = v_{\rm B-L} =
O(0.1)\,M_*$.  However, we can suppress the above interaction by
assigning the $R$-charges as $R[\Phi{\bar \Phi}] = R[H_1 H_2] = 0$. If
the $R$-symmetry is dominantly broken by the constant term in the
superpotential, the coefficient $c$ is expected to be of
$O(m_{3/2}M_P^2/M_*^3)$, and such a coupling becomes irrelevant.\footnote{
  There can be an operator, $W= (\Phi \bar{\Phi}) T Q H_2/M_*^2$,
  where $T$ and $Q$ denote the right-handed quark and left-handed
  quark doublet in the third generation, respectively. The decay rate
  through this operator is suppressed by $O(10^{-2} v_{\rm B-L}^2
  M^2/M_*^4)$, compared to that into a lepton and slepton pair, and
  therefore, does not change our arguments.
}

Lastly we emphasize that the longevity of the $\lambda_1$ DM arises
from the geometrical separation and the hierarchy between $M$ and $m$,
not from conservation of some discrete symmetry such as the
$R$-parity. In fact, we assume that in our scenario the lightest
neutralino is the lightest SUSY particle (LSP) and it is absolutely
stable with the conserved $R$-parity. Therefore, in our set-up, there
are actually two DM candidates, $\lambda_1$ and the neutralino LSP. We
assume that $\lambda_1$ is the dominant component of the DM for the
moment, and will mention such a case that the $\lambda_1$ is
subdominant in Sec.~\ref{sec:4}

\section{Electron and positron excesses from the decaying hidden-gaugino DM}
\label{sec:3}
As we have shown in the last section, the $\lambda_1$ is almost stable
but has some decay modes through a small mixing with the $\lambda_{\rm B-L}$.  If the DM is
heavier than the sleptons, the DM can decay into slepton + lepton and
this slepton causes the SUSY cascade decay and reaches the LSP,
emitting high energy SM particles.  In the case of the slepton NLSP,
the slepton emits only SM leptons.  The anomalies observed in the recent
$e^{\pm}$ cosmic-ray experiments can be explained by the energetic
leptons from this DM decay.  In addition, if the DM mass $m$ is
lighter than the mass of squarks, the DM causes almost no hadronic
decay which would produce many antiprotons and
photons~\cite{Arvanitaki:2008hq}.\footnote{This assumption seems to be
  natural, because usual SUSY breaking model indicate that the squarks
  are much heavier than (right-handed) slepton.  In addition, the
  SM-like lightest higgs mass bound
  ($m_{h^0}>114.4$~GeV~\cite{Barate:2003sz}) also indicate heavy
  squarks ($\gsim 1$ TeV.)  (The mass of SM-like higgs receives a
  radiative correction from quark-squark loop diagrams.) }

Now let us estimate the cosmic ray signals from the DM decay.  To
estimate the energy spectrum of the decay products of the DM, we have
used the program PYTHIA~\cite{Sjostrand:2006za}.  The particles
produced in the DM decays are influenced by various factors in the
propagation.  For the propagation in the Galaxy, we adopt the method
discussed in Refs.~\cite{Ibarra,Hisano:2005ec} with the Navarro, Frenk
and White halo profile~\cite{Navarro:1996gj};
\beq
\rho_{DM}=\frac{\rho_0}{(r/r_c)[1+(r/r_c)]^2},\label{eq:NFW}
\eeq
where $\rho_0=0.26~ {\rm GeV cm^{-3}}$ and $r_c= 20~{\rm kpc}$.  As a
diffusion model for the electron and positron cosmic rays, we use the
MED model in Ref.~\cite{Delahaye:2007fr}.  As for the gamma ray
signal, we have averaged the halo signal over the whole sky excluding
the region within $\pm 10^\circ$ around the Galactic plane.  For the
extragalactic component, the gamma ray is influenced by the red-shift.
We estimate the extragalactic component by using the following
cosmological parameters; $\Omega_{\Psi}h^2\simeq0.11,~\Omega_{\rm
  matter}h^2\simeq0.13,~\Omega_{\Lambda}\simeq0.74,~\rho_c \simeq
1.0537\times 10^{-5} h^2~ {\rm GeV cm^{-3}},~h\simeq
0.72$~\cite{Komatsu:2008hk}.

\begin{figure}[htbp]
\begin{tabular}{cc}
\begin{minipage}{0.5\hsize}
\begin{center}
\epsfig{file=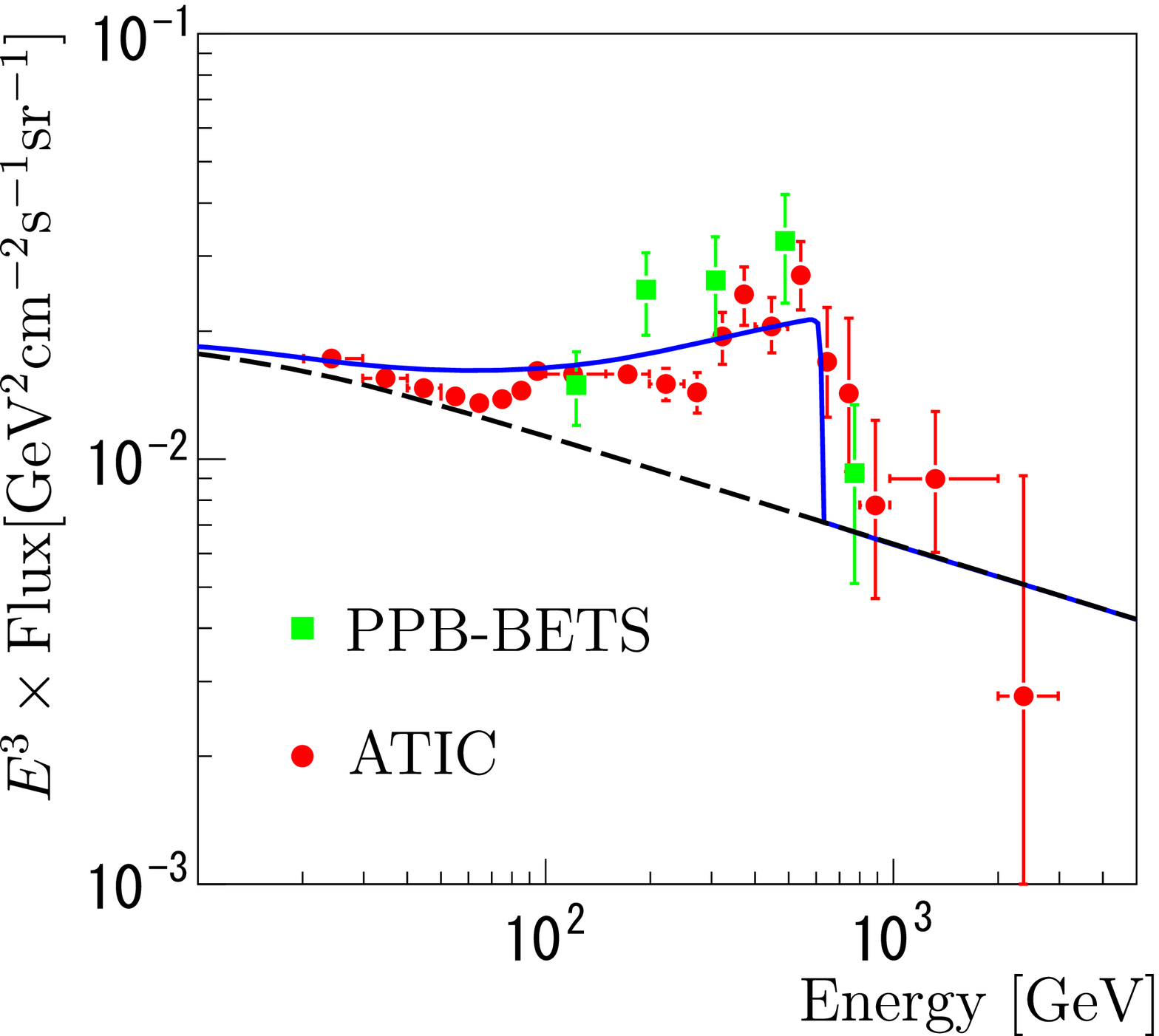 ,scale=.42,clip}\\
(a)
\end{center}
\end{minipage}
\begin{minipage}{0.5\hsize}
\begin{center}
\epsfig{file=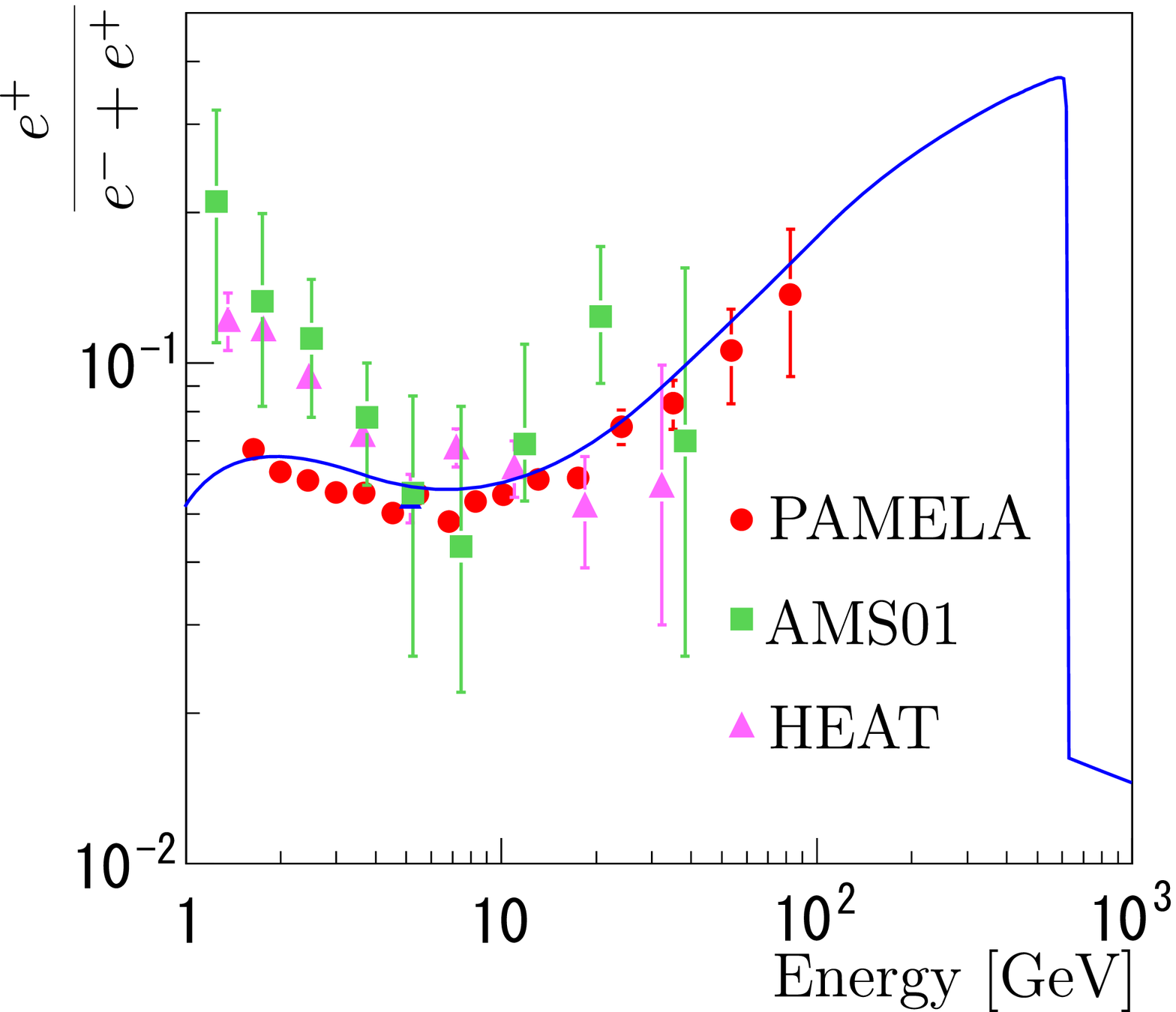 ,scale=.42,clip}\\
(b)
\end{center}
\end{minipage}\\\\
\multicolumn{2}{c}{
\begin{minipage}{1\hsize}
\begin{center}
\epsfig{file=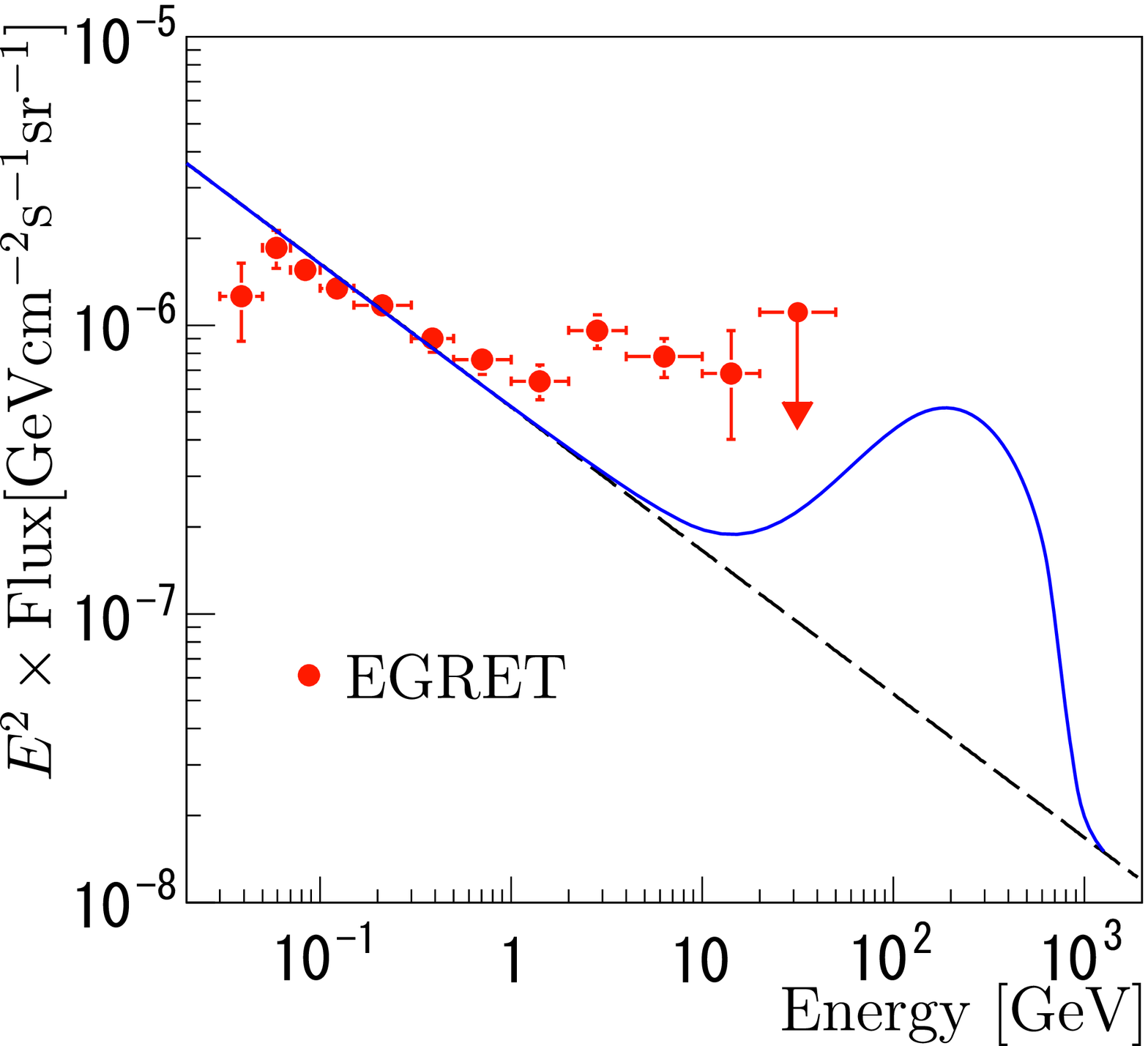 ,scale=.42,clip}\\
(c)
\end{center}
\end{minipage}}
\end{tabular}
\caption{Cosmic ray signals in the present model.
(a): positron and electron fluxes with experimental data~\cite{:2008zz,Torii:2008xu}.
(b): positron fraction with experimental data \cite{Adriani:2008zr,Aguilar:2007yf,Barwick:1997ig}.
(c): gamma ray flux with experimental data~\cite{Sreekumar:1997un,Strong:2004ry}.
}
\label{fig:signal}
\end{figure}

We set the DM mass equal to $1300$ GeV with lifetime $8\times 10^{25}$
sec.  As for the SSM spectrum, we assume that the LSP is the bino-like
neutralino of a mass $148$ GeV and the NLSPs are selectron, smuon, and
stau with a mass $150$ GeV. Note that, with our choice of the mass
parameters, thermal relic abundance of the neutralino LSP is
suppressed due to efficient co-annihilation; this is to make sure that
the neutralino LSP abundance is negligibly small compared to that of
$\lambda_1$. (As for the production of $\lambda_1$, see the next section.)  
For simplicity, we assume that the other SM scalar
partners are heavier than  the $\lambda_1$ and that the  $\lambda_1$  decays only into the
the slepton $+$ lepton.  In Fig.~\ref{fig:signal}, we show the cosmic ray signals.  As
for the electron and positron background, we have used the estimation
given in Refs.~\cite{Moskalenko:1997gh,Baltz:1998xv}, with a
normalization factor $k_{\rm bg} = 0.7$.  We have taken into
consideration the solar modulation effect in the current solar
cycle~\cite{Baltz:1998xv}.  We set the gamma ray background flux as
$5.18\times 10^{-7}(E/1~\GEV)^{-2.499}~{\rm GeV^{-1}
  cm^{-2}s^{-1}sr^{-1}}$ as in Ref.~\cite{Ishiwata:2008cu}.  We have
assumed the energy resolution is $15 \%$ for the gamma-ray signal.  As for electron and
positron signal, this model can explain the cosmic ray anomalies quite
well.  The gamma ray signal is not in conflict with the currently
available experimental data.  Furthermore, our model predicts a bump
around several hundred GeV in the diffuse gamma-ray spectrum.  Such a
feature in the gamma-ray spectrum can be tested by the Fermi
satellite~\cite{FGST} in operation.

\section{Discussion and conclusions}
\label{sec:4}

Let us here discuss the production mechanisms of the hidden gaugino,
$\lambda_1$, in the early universe.  As we have seen in the previous sections, the
interactions of the $\lambda_1$ with the SSM particles are extremely
suppressed due to both the geometrical separation and a large
hierarchy between $M$ and $m$.  In order to produce a right amount of
the $\lambda_1$ DM from particle scatterings in thermal plasma, the
reheating temperature after inflation has to be close to the $B-L$
breaking scale. Such a high reheating temperature is in conflict with
the constraint from the notorious gravitino
problem~\cite{Kawasaki:2008qe}.\footnote{
  Note that the gravitino must be heavier than the hidden gaugino,
  since the hidden gaugino will decay into the gravitino and the
  hidden gauge boson with a much shorter lifetime, otherwise.
} It is therefore difficult to produce $\lambda_1$ by thermal
scatterings.  Nevertheless, the right abundance of the hidden gaugino
can be non-thermally produced by the inflaton decay, if the inflaton
has a direct coupling to the hidden gauge sector.

The gravitino decay is another possible way to produce
$\lambda_1$. Due to the number of degrees of freedom, the gravitino
will mainly decay into the SSM sector, and the branching ratio of the
the hidden gaugino production is expected to be $O(1)$\%. Therefore,
the hidden gaugino can only be a subdominant component of the total DM
in this case.  Taking account of the presence of the stable neutralino
LSP, however, this is not a problem. For instance, for the reheating
temperature $T_R \sim 10^6$ GeV, the abundance of the hidden gaugino
produced from the gravitino decay is estimated to be $\Omega_{\lambda}
h^2 = 10^{-6} - 10^{-5}$.  The hidden gaugino decay can still account
for the cosmic-ray anomalies if we adopt a slightly smaller value of
$M \sim 10^{15}$ GeV.

With more than two extra spatial dimensions, the SUSY breaking sector
may reside on another brane, and the anomaly-mediation may be
realized.  The presence of light sleptons can then be naturally realized, and
the LSP is likely the wino. Due to its large annihilation cross section,
the thermal relic abundance of the wino will be much smaller than the observed DM abundance.
Then it will become unnecessary to assume the stau neutralino co-annihilation 
in order to suppress the neutralino LSP abundance, 
and thus, the slepton masses are not necessarily tied to the LSP mass. 
As for the production of $\lambda_1$,
it is possible that the
inflaton decay produces a right amount of  $\lambda_1$ through either direct or
anomaly-induced couplings~\cite{Endo:2007ih}.\footnote{
  For the anomaly-induced decay to proceed, we need to introduce
  hidden matter fields charged under the hidden U(1) gauge
  symmetry~\cite{Endo:2007ih}.
} 
Since the gravitino mass can be as heavy as $100$\,TeV,
the gravitino problem is greatly relaxed.
The gravitino decay can also produce the hidden gaugino
as well as the wino, and the fraction of the $\lambda_1$ in the total DM can be as
large as $O(1)$\% in this case, while most of the total DM is the wino non-thermally
produced by the gravitino decay.

We have so far assumed that the gauge symmetry in the bulk is
U(1)$_{\rm B-L}$, but it is possible to consider another anomaly-free
symmetry given by a linear combination of the U(1)$_{\rm B-L}$ and the
hypercharge U(1)$_{\rm Y}$. For instance, if the gauge symmetry in the
bulk is identified with a U(1)$_5$, so-called
``fiveness"~\cite{Fujii:2002mb}, the hidden gaugino is coupled to the
higgs and higgsino, in addition to lepton + slepton and quark +
squark.  However, if the higgsino mass is heavier than $m$, the hidden
gaugino still mainly decays into lepton + slepton, thereby keeping the
suppression of the antiproton production.

The decay into a quark pair and the SSM gauginos can actually occur
through a virtual squark exchange,
with a suppressed rate. Compared to the main decay modes into a lepton
and slepton pair, the partial decay rate is suppressed by a factor of
$O(10^{-3})$ for the squark mass of a few TeV. Such a suppression is
small enough to make the DM contribution to the antiproton flux
consistent with the observation.

In this paper we have proposed a scenario that a hidden U(1) gaugino
constitutes the DM and decays mainly into the leptons through a mixing
with a U(1)$_{\rm B-L}$. We have shown that the energetic leptons from
the DM decay can account for the recently reported PAMELA and ATIC
anomalies in the cosmic-ray electrons/positrons.  We should emphasize
that our model is free from an overproduction of antiprotons, if the
squarks are heavier than $m \sim 1.3$ TeV. The predicted excess in the
diffuse gamma-ray flux around several hundred GeV can be tested by the
Fermi satellite.

\section*{Acknowledgement}
One of the authors (FT) was aware that there is another group
working independently on a scenario using a hidden gaugino~\cite{IETW}.
This work was supported by World Premier International Center
Initiative (WPI Program), MEXT, Japan.  The work of SS is supported in
part by JSPS Research Fellowships for Young Scientists.

\end{document}